\documentclass[aps,pre,superscriptaddress,twocolumn,floatfix]{revtex4-1}
\usepackage{graphicx}
\usepackage{dcolumn}
\usepackage{bm}
\usepackage{hyperref}
\usepackage{amssymb}
\usepackage{amsmath}
\usepackage{array}
\usepackage[normalem]{ulem}

\usepackage{rotating} 

\usepackage{siunitx}

\usepackage{natbib}
\usepackage{epstopdf}
\usepackage{color}
\usepackage{makecell}
\usepackage[table]{xcolor}
\usepackage{tabularx}
\usepackage{ragged2e}

\usepackage{lmodern}
\usepackage[latin1]{inputenc} 

\graphicspath{{figure/}} 

\definecolor{lightgray}{gray}{0.9}
\definecolor{darkgray}{gray}{0.60}
\newcolumntype{C}{c<{\kern\tabcolsep}<{\kern\tabcolsep}@{}}
\begin{document}
\clearpage

\newcommand{\be}{\begin{equation}}
\newcommand{\ee}{\end{equation}}
\newcommand{\del}{\partial}

\let\oldAA\AA
\renewcommand{\AA}{\text{\normalfont\oldAA}}

\newcommand{\LL}[1]{\textcolor{red}{{\bf LL:} #1}}
\newcommand{\MT}[1]{\textcolor{red}{{\bf MT:} #1}}
\newcommand{\Ch}[1]{\textcolor{cyan}{{\bf Ch:} #1}}
\newcommand{\LH}[1]{\textcolor{blue}{{\bf LH:} #1}}
\newcommand{\JFJ}[1]{\textcolor{purple}{{\bf JFJ:} #1}}
\newcommand{\PBRO}[1]{\textcolor{cyan}{{\bf PBRO:} #1}}

\preprint{APS/123-QED}

\newcommand{\udem}{D\'{e}partement de Physique and Regroupement
 Qu\'{e}b\'{e}cois sur les Mat\'{e}riaux de Pointe, Universit\'{e} 
de Montr\'{e}al, C.P. 6128, Succursale Centre-Ville, Montr\'{e}al,
 Qu\'{e}bec, Canada H3C~3J7}
\newcommand{\nrc}{National Research Council of Canada, Ottawa, Ontario,
 Canada K1A 0R6}

\title{
Pair potentials for warm dense matter and their application
to x-ray Thomson scattering in aluminum and beryllium
}

\author{L. Harbour}\email[Email address:\ ]{louis.harbour@umontreal.ca}\affiliation{\udem}
\author{M. W. C. Dharma-wardana}\email[Email address:\ ]{chandre.dharma-wardana@nrc-cnrc.gc.ca}\affiliation{\nrc}
\author{D. D. Klug}\affiliation{\nrc}
\author{L. J. Lewis}\affiliation{\udem}

\date{\today}

\begin{abstract} 
Ultrafast laser experiments yield increasingly reliable data on warm dense
matter, but their interpretation requires theoretical models. We employ an
efficient density functional neutral-pseudoatom hypernetted-chain (NPA-HNC) 
model  with accuracy comparable to \textit{ab initio}
 simulations and which provides 
first-principles pseudopotentials and pair-potentials for warm-dense matter. It avoids the use of 
(i) \textit{ad hoc} core-repulsion models and (ii) ``Yukawa
screening'', and (iii) need not assume ion-electron thermal equilibrium. 
Computations of the x-Ray Thomson scattering (XRTS) spectra of aluminum
 and beryllium are compared with recent experiments  and with
density-functional-theory molecular-dynamics (DFT-MD) simulations. 
The NPA-HNC structure factors, compressibilities,
phonons and conductivities agree closely with DFT-MD results, while Yukawa screening
gives misleading results. The analysis of the XRTS data
for two of the experiments, using two-temperature quasi-equilibrium models,
is supported by calculations of their temperature relaxation times.

\end{abstract}

\maketitle

\section{Introduction}

High-energy deposition on matter using ultrafast lasers has opened the way to novel
non-equilibrium regimes of density and temperature, raising issues of broad
scientific interest~\cite{WDM-book}. These include hollow atoms, quasi-equilibrium solids
 and transient plasmas. The physics of 
warm dense matter (WDM) applies to hot carriers in nanostructures, space
reentry,  inertial confinement
fusion~\cite{Dimonte08,ICF-Graz11}, Coulomb explosion, laser machining, surface
ablation~\cite{Lewis-Abl-03a, Lewis-Abl-03b,Lewis-Abl-06} and astrophysical environments,
etc. The interactions in the WDM regime are  characterized by the effective
coupling parameter $\Gamma$, viz.,  the ratio of the Coulomb energy to the kinetic
energy, which is bigger than unity. Simple approaches based on
perturbation theory from a known ``ideal'' state thus become inapplicable.

Recent laser experiments on solid simple metals have reached WDM conditions
through  e.g., (i) ultrafast isochoric heating ($\rho =\rho_0$, where $\rho_0$
$\rho$ are the initial and final densities, respectively)~\cite{Milsch-88,chen2013,Sper15},
and (ii) shock-compression 
($\rho>\rho_0$)~\cite{Flet-Al-15,Saiz-Li-08, Ma-Al-13,Lee-Be-09,Glenzer-Be-07}. 
In situation (i) the optical laser directly interacts with a metallic target
 and couples to the free electrons causing their
temperature $T_e$ to reach many eV, while ions remain approximately at their
initial temperature $T_i$. In situation (ii) the laser may pre-couple
 to the covalent electrons (bonds)  of a non-metallic driver layer placed prior
 to the target material. This sets up a
shock wave that can both heat and compress the target material which is usually
metallic. If the driver layer is thick enough, the $T_i$ attained by the
target exceeds $T_e$ as the shock wave does not directly couple to
the electrons. A  third and more complex situation (iii) arises if the insulating driver
layer creates a shock wave as before, but in addition the laser penetrates through it and
deposits energy directly in the metallic target layer. The electron 
temperature $T_e$ can then exceed the ion temperature $T_i$ even in shock-compression
experiments. Finally, the state of the WDM encountered by the probe beam
 also depends on the time delay $\tau_d$ between the pump
laser and the probe laser~\cite{chen2013,elr2001}. If $\tau_d$ significantly
exceeds the electron-ion temperature relaxation time $\tau_{ei}$, $T_e$ and
$T_i$ would then relax to a common equilibrium temperature $T$. It should be
noted that as $T_e$ approaches $T_i$, the temperature relaxation becomes
increasingly slower, and coupled-mode formation begins (on phonon timescales) and
the process is further slowed down~\cite{elr2001}. Hence the tacit assumption of
thermal equilibrium in WDM created by laser-shock techniques can produce
misleading interpretations of experiments, as we show in what follows. 

In the discussion above we have assumed the simplest non-equilibrium paradigm,
viz., the well-known two-temperature ($2T$) model~\cite{Anisimov74}. However,
this may be too simplistic. The laser may create spatial and thermal
inhomogeneous distributions which are hard to interpret.  On  short timescales
(e.g., $<$100 fs) or in more complex situations, even the electrons may  not
equilibrate to a common, unique temperature $T_e$~\cite{Medvedev}.

X-ray Thomson scattering  (XRTS) is a key method for studying WDM
as it yields  $T_e$, $T_i$, the ion density $\rho$, the mean electron density
$n_e$,  and details of ionic and electronic correlations. The XRTS signal is
directly proportional to the total electron-electron dynamic structure factor
$S_{ee}(k,\omega)$, which naturally follows a decomposition in terms of free-free,
bound-bound  and bound-free  contributions from all ``{\it single ion sites}'', as discussed
by Chihara~\cite{Chihara}. Such a decomposition is not available \textit{directly} via
density-functional theory (DFT) calculations, which use an $N$-ion simulation
cell, since the electron density $n(r)$ calculated by such methods is the
property of all the $N$ ions. However, by combining DFT with molecular-dynamics (MD) simulations (DFT-MD), 
the known ionic positions permit the calculation of the static ion-ion, ion-electron 
structure factors and the electron density at a  ``{\it single ionic center}''. 

The work of Vorberger {\it et al.}\cite{Vorberger15} demonstrates the interest in  simpler methods to obtain such `single-ion' 
properties as charge densities $n(q)$ for WDM studies. In Ref.~\cite{
Vorberger15} $n(q)$ is calculated from an {\it externally obtained}
pseudopotential (for Al). Since such potentials are not available for WDM 
conditions, the authors
use an Ashcroft empty-core potential $U^{emp}(q)$~\cite{Ashcroft63,Ashcroft66}.
However,  such  $U^{emp}(q)$ are applicable only for a few  metals like Al at 
normal density and temperatures. The advantages and shortcomings of the 
empty-core pseudopotential even for aluminum at normal density and temperature are well 
known, and more complex transferable pseudopotentials are used in DFT-MD codes. 
The alternative of deconvoluting the $N$-ion charge density obtained from DFT-MD 
into a single-center charge density is a computationally demanding complex 
process.  Instead,  the neutral-pseudoatom (NPA)  method directly constructs 
density- and temperature-dependent pseudopotentials {\it in situ} (without using 
transferable potentials) via an all-electron calculation. It is a rigorous DFT 
formulation that uses an {\it effective} single-ion model
of the electron-ion system to provide all the required quantities directly in order to predict, say, the XRTS signal, with negligible computational cost. The only term that depends directly on $T_i$ is the contribution to XRTS from the isotropic \textit{ion feature} $W(k,\omega)$, sometimes referred to as the   Rayleigh  feature (see e.g.  Refs.~\cite{GlenzerRMP,Gregori06});
it is given by
\begin{eqnarray}
\label{xrts-eqn}
W(k,\omega) &=& |f(k)+ 
q(k)|^2 S_{ii}(k,\omega),\\
S_{ii}(k,\omega) &\simeq& S_{ii}
(k)\delta(\omega).
\end{eqnarray}
Here $f(k)$ and $q(k)$ are the form factors of bound $n_b(r)$ and free $n_f(r)$
electron densities {\it at an individual ion}, and  $S_{ii}(k,\omega)$ is the
dynamic structure factor of the ions. Current XRTS experiments cannot  resolve
ion dynamics (at meV energy scales); hence it  is  approximated by the static
structure factor  $S_{ii}(k)$, denoted hereafter as $S(k)$. Thus, while $T_e$
and $n_e$ are determined via the inelastic part of the XRTS signal, a
determination of $T_i$ is required to obtain the ion-ion $S(k)$.

Such computations of the XRTS signal have mostly been done with electronic-structure
codes~\cite{VASP,Abinit} based on DFT Kohn-Sham calculations for a fixed
set of $N$ ions held in a simulation box, coupled with MD
to move them and generate ensemble averages for observable properties. Results
from these computationally intensive DFT-MD simulations are themselves fitted
 to intermediate
quantities, e.g., simple ``physically-motivated'' pair potentials, to ease
computations. Such  intuitive models usually have hidden  pitfalls but become
entrenched as accepted practice unless corrected. 

The objective of the present study is to employ
the  DFT-based NPA approach to provide simple
first-principles calculations of the electron densities, $n(k)$,
 pseudo-potentials, $U_{ei}(k)$, and ion-ion
pair interactions, $V_{ii}(k)$. Here, by `first-principles' we mean
calculations that do not recourse to {\it ad hoc} intermediate models, but use
only results flowing from the initial Kohn-Sham Hamiltonian of the NPA formulation. Admittedly, in calculating $S_{ii}(k)$ using
an integral equation, a hard-sphere
bridge parameter $\eta$ is invoked. But it is determined by an 
optimization procedure internal to the method; or
it may be avoided altogether by using MD with the NPA pair-potentials, as discussed below.

The pair-potentials  when coupled with a
 hypernetted-chain (HNC) integral equation or MD yield
  structure factors $S(k)$  which can be use to calculate all
other physical properties of WDM  when used with
the pseudopotentials and charge densities. In particular, all
quantities needed for computing XRTS spectra, transport properties,
energy relaxation, equation of state (EOS)  etc., become available and may be used
to investigate recent experiments as well as the quality of
their interpretations employing popular
``physically-motivated'' {\it ad hoc} models. Since the NPA-HNC method is typically as accurate as
DFT-MD, while orders of magnitude more efficient, it permits the rapid computation of
$W(k)$ for several $T_i$ in a very effective manner, enabling us to examine different
$2T$-models and their consistence with experiment. 

In particular, the need for a simple 
potential has led to the use of  an
intuitive model that has come into vogue with WDM studies, viz.\
the  ``Yukawa + short-ranged repulsive (YSRR) potential''~\cite{Flet-Al-15,Ma-Al-13,Ye11},
\begin{equation}
 \beta_i V_{ii}^{\text{ysrr}}(r) =
\sigma^4/r^4+\beta_i\exp(k_sr)/r,
\end{equation}
 introduced in Ref.~\cite{Wunsch09}.
Here, $\beta_i = 1/T_i$ is the inverse ion temperature, $k_s$ is a screening wavevector
and $\sigma$ is a parameter fitted to MD data.  We examine the validity of the YSRR
approach using first-principles models and XRTS data for $2T$ systems  ($T_e\ne
T_i$) as well as for equilibrium systems. The YSRR potential is found to yield 
misleading conclusions about $T_e$ and $T_i$, incorrect compressibilities  (i.e.,
a property of the EOS), incorrect phonons and incorrect electrical
conductivities.

The  utility of the NPA-HNC and the possibility of two-temperature systems in
 laser-generated WDM is illustrated below by re-analyzing experiments on Al and Be.
 For Al, we examine shock-compressed systems by (i) Fletcher 
\textit{et al.}~\cite{Flet-Al-15}  at $\rho/\rho_0 = 2.32$  and $T_e =
 1.75$ eV and by (ii) Ma \textit{et al.}~\cite{Ma-Al-13} 
 at $\rho/\rho_0 = 3.0$ and $T_e = 10$ eV with $\rho_0=$ 2.7 g/cm$^3$; for Be,
 we examine the shock-compressed system by (iii) Lee 
\textit{et al.}~\cite{Lee-Be-09} at $\rho/\rho_0 = 2.99$ and $T_e = 13$
 eV and the isochorically-heated system by (iv) 
Glenzer \textit{et al.}~\cite{Glenzer-Be-07} at $\rho/\rho_0 = 1$ and
 $T_e = 12$ eV with $\rho_0=$ 1.85 g/cm$^3$.
\section{The NPA-HNC model}
An XRTS $W(k)$ calculation  needs the electron density
 at an ion and
the structure $S(k)$  of the system. The NPA
approach~\cite{Dagens1,Dagens2,DWP1,NPA-PDW} decomposes the total 
density into a superposition of effective one-center  densities combined
{\it via}
structure factors  and provides a comprehensive scheme based on
DFT. However, this is not intrinsically a superposition {\it approximation}; rather,
this is a rigorous method in DFT which is often not recognized as such,
 with a tendency to consider it as a mean-field average-atom model. In effect,
 DFT provides a route to an {\it exact} average-atom description of an arbitrary
 electron-ion system. As discussed in Ref.~\cite{DWP1}, DFT asserts that the
free energy $F[n,\rho]$ is a functional of the one-body electron density
 $n(\mathbf{r})$, and the one-body ion density $\rho(\mathbf{r})$, irrespective
 of the existence of complex interactions (e.g., superconductive associations
 for electrons), and complex covalent-bonding structure, $d$-bonds etc.,
 for ions. Furthermore, the functional derivatives of $F[n,\rho]$ satisfy
 the following stationary conditions:
\begin{eqnarray}
\label{ks-el.eq}
\delta F[n,\rho]/\delta n &=&0, \\
\label{ion-dft.eq}
\delta F[n,\rho]/\delta \rho&=&0.
\end{eqnarray}  
Standard DFT uses the first stationary condition to construct a Kohn-Sham
one-body potential acting on an effective ``one-electron'' density $n(\mathbf{r})$.
Similarly the stationary condition on the ion density,
Eq.~\ref{ion-dft.eq},  defines a set of non-interacting ``Kohn-Sham ions''
 moving in the classical form of the DFT potential acting on a
  ``single ion'' representative of $\rho(\mathbf{r})$. The ions
 can be regarded as classical spinless particles for our
purposes. In Ref.~\cite{DWP1} the potential acting on the ``average ion''
 was identified with the ``potential of
mean-force'' used in the theory of classical liquids \cite{Hansen}.
 Such an approach requires corrections beyond mean-field theory which are included
in the exchange-correlation functional $F^{ee}_{xc}[n]$ for the electrons {\it
and} in the ion-ion correlation functional  $F^{ii}_{c}[\rho]$ for the ions.
The ions are classical and do not have exchange (see Eq.~1.13,
in Ref.~\cite{ilciacco95}). Only pair-interactions between the DFT average-ions
appear in the theory and the burden of approximating effects beyond pair-interactions
falls on constructing these correlation functionals. It has in fact
been shown that  such an approach is successful even for liquid carbon with
 transient multi-center covalent bonding where carbon-carbon interactions
 are usually handled with multi-center
  ``reactive bond-order potentials''~\cite{Lqd-C2016}.

However, here we study Al and Be in regimes where they are
expected to be `simple liquids'. A sum of hypernetted-chain diagrams and bridge
 diagrams has been used to model $F^{ii}_{c}[\rho]$.
The usual DFT-MD codes  do not use Eq.~\ref{ion-dft.eq} nor
 $F^{ii}_c$ and $F^{ei}$ since the ion many-body problem
 is not reduced to a one-body problem for the ions,
unlike in the NPA. Instead, standard DFT uses a Born-Oppenheimer
approximation where $N$ ions in a simulation box  are explicitly enumerated and
provide an ``external potential'' to the Kohn-Sham electrons. In contrast,
the NPA uses only the one-body ion distribution $\rho(\mathbf{r})$.
 Given a good ion-correlation functional $F_c^{ii}(\rho)$,
 enormous computational
simplifications follow from this full-DFT approach compared to the standard
method which calculates the Kohn-Sham eigenfunctions of a  simulation
cell containing typically $N$ = 64 - 128 ions. Thus the ``single-center'' NPA is a
{\it rigorous} DFT average-atom approach, and its approximations lie in the
construction of $F_{c}^{ii}$ and $F_{xc}^{ee}$. The other advantage is that
the NPA naturally provides
``single-ion'' properties like the mean ionization $\bar{Z}$, charge
density $n(\mathbf{r})$ at a single ion, and the separation of the bound-electron and
 free-electron spectra (needed in XRTS theory) consistent with the
 exchange and correlation potentials used in the theory.

Several NPA models are described in the literature~\cite{Sandia1}, e.g.,
those using ion-sphere models and other  prescriptions not completely based
on DFT theory. These different formulations affect how the chemical
potential   is treated and how the bound and free electrons are
identified~\cite{NPA-PDW,NPA-other1,NPA-other2,NPA-other3,NPA-other4,NPA-other5}. 
We employ the NPA model of Perrot and
Dharma-wardana~\cite{NPA-PDW,NPA2, CPP-cdw} which includes a cavity of radius $r_{ws}$, 
with $r_{ws}=\{3/(4\pi\rho)\}^{1/3}$ the ion Wigner-Seitz radius,
around the central nucleus to mimic, in a  simplified way, the
 ion-density $\rho(r)$ of
the plasma contained in a  ``correlation sphere'' of radius $R_c \sim 10
r_{ws}$. This is equivalent to using $(4/3)\pi (R_c/r_{ws})^3$, i.e.,
about 4200 particles in an MD simulations; in contrast, typical DFT-MD simulations use about 250 particles.
  The full ion distribution is subsequently evaluated by an HNC
 or modified-HNC (MHNC)
procedure although MD may also be used, especially if low-symmetry situations
are envisaged. The $R_c$ is such that the pair-distribution functions (PDF),
viz., ion-ion $g_{ii}(r)$ and
 ion-electron $g_{ie}(r)$ have asymptotically
reached unity as $r\to R_c $. The electron-electron PDF
$g_{ee}$ can be shown to also reach the asymptotic limit when $r\to R_c$ as the e-e coupling
is comparatively much weaker. The electron chemical potential is for 
non-interacting electrons at the interacting mean density $n_e$  and temperature
$T_e$, as required by DFT.  The finite-$T$ DFT calculations are done using  a
finite-$T$  exchange-correlation functional $F_{xc}^{ee}$~\cite{PDWXC}. 
The free-electron density $n_f(r)$ is calculated using
 Mermin-Kohn-Sham wave functions
which are orthogonal to the  core states. Core-valence Pauli blocking,
core-repulsions as well as core-continuum exchange-correlation effects are
naturally included in the model. The NPA is an all-electron calculation
 and yields bound-state energies, bound-electron densities, as 
well as continuum densities
and phase shifts which satisfy the Friedel sum rule.

The NPA  free-electron pileup $n_f(k)$ around the NPA-nucleus is the key
quantity in constructing electron-ion pseudopotentials $U_{ei}(k)$ and
 ion-ion pair potentials $V_{ii}(k)$, given in terms of the fully interacting
static electron response function $\chi(k,T_e)$ as follows :
\begin{eqnarray}
\label{pseudo.eq}
U_{ei}(k) &=& n_f(k)/\chi(k,T_e),\\
\label{resp.eq}
\chi(k,T_e)&=&\frac{\chi_0(k,T_e)}{1-V_k(1-G_k)\chi_0(k,T_e)},\\
\label{lfc.eq}
G_k &=& (1-\kappa_0/\kappa)(k/k_\text{TF});\quad V_k =4\pi/k^2,\\
\label{ktf.eq}
k_{\text{TF}}&=&\{4/(\pi \alpha r_s)\}^{1/2};\quad \alpha=(4/9\pi)^{1/3},\\
\label{vii.eq}
 V_{ii}(k) &=& \bar{Z}^2V_k + |U_{ei}(k)|^2\chi_{ee}(k,T_e).
\end{eqnarray}
Here $\chi_0$ is the finite-$T$ Lindhard function, $V_k$ is the bare
Coulomb potential and $G_k$ is a local-field correction (LFC).
Hence the electron response  goes beyond the
random-phase approximation (RPA). The finite-$T$ compressibility sum rule is
satisfied since $\kappa_0$ and $\kappa$ are the non-interacting and
interacting electron compressibility, respectively, with  $\kappa$ matched to the
$F_{xc}$ used in the Kohn-Sham calculation. In Eq.~\ref{ktf.eq}, $k_\text{TF}$
appearing in the LFC is the Thomas-Fermi wavevector. We use a $G_k$ evaluated at
$k\to 0$ for all $k$ instead of the more general form (e.g., Eq.~50  in
Ref.~\cite{PDWXC}) since the $k$-dispersion in $G_k$ has negligible effect for
the WDMs treated in this study. Note that the ``Yukawa form'' of the
pair-potential is obtained from the above equations at sufficiently
 high temperatures since the Lindhard function can be approximated by
its $k\to 0$ limit under such Debye-H\"{u}ckel-like conditions,
 while $G_k$ goes to zero. Such
approximations are largely invalid in the WDM regime; 
Friedel oscillations in the pair-potentials contribute to defining
 the peak positions in the $g(r)$ and hence their relevance to observed properties is well-known experimentally and theoretically.
Furthermore the need for finite-$k$ screening instead of the Yukawa form 
is the norm for systems with $T/E_F<1$ and normal densities.
In fact the pair-potentials, PDFs, XRT features, conductivities and  phonons will be incorrect if calculated from the $k\to 0$ Yukawa form for the given conditions. Hence all the
observable properties studied here can be regarded
as examples of observations of finite-$k$ screening 
(see, e.g.,  Ref~\cite{Chapman15}).

The pseudopotential $U_{ei}(k)$ is a {\it local} potential which
contains  non-linear effects as the $n_f(k)$ was calculated from the
Kohn-Sham equations. 
However, since it is forced to be linear in the response,  the
pair-potential Eq.~\ref{vii.eq} is trivially constructed. The regime
of validity of this procedure is discussed in Ref.~\cite{2Tpp}. Outside the regime of validity it becomes increasingly more approximate, but as 
no {\it ad hoc} models extraneous to the calculation are invoked, it remains a first-principles method for
constructing the pseudopotential from the all-electron single-center Kohn-Sham calculation.
  The
structure factor $S(k)$ is computed (for uniform systems) from the modified
hypernetted-chain (MHNC) equation which includes a bridge term $B(\eta,r)$
modeled using a Percus-Yevik hard-sphere fluid with a packing fraction
$\eta$. The packing fraction is determined by the Lado-Foiles-Ashcroft
  {\em et al.} (LFA)
criterion~\cite{LFA, chenlai} based on the Gibbs-Bogoluibov inequality.
Although we use the MHNC equation, we refer
to the general method as NPA-HNC, or occasionally as NPA-MHNC when we wish
to emphasize the use of the MHNC procedure over the HNC one. It should also
 be noted that any ambiguity 
in the choice of the bridge function,
or the use of a hard-sphere model for the bridge
function, can be avoided if the NPA pair-potential
is used directly in MD to generate $g(r)$. Such NPA+MD calculations were not deemed
 necessary  in the present study.
 
Since the NPA pair-potential accurately predicts
phonons (i.e., meV scales of energy) for common $2T$ WDM systems~\cite{CPP-Harb},
even the dynamical $S_{ii}(\mathbf{k},\omega)$ can be predicted when  XRTS
data at meV accuracy become available. Furthermore, since all the PDFs and
interaction potentials are available, the Helmholtz free energy $F$, and hence
EOS properties, specific heats etc.,  as well as linear transport properties,
can be calculated rapidly and in a parameter-free manner. Many such
 calculations have been presented in the past, as reviewed
 in Ref.~\cite{CPP-cdw}. Here we
illustrate this for XRTS experiments on
an equilibrium system and on a $2T$-quasi-equilibrium system.
%
%
\section{Aluminum}
\subsection{ Shock-compressed Aluminum - I}
Using XRTS, Fletcher {\em et al.}~\cite{Flet-Al-15}  have studied compressed
aluminum evolving across the melting line  into  a WDM state (named Al-I hereafter).
 From their inelastic data, they determined the aluminum density and temperature 
to be $\rho/\rho_0 = 2.32$  and $T_e = 1.75$ eV, respectively. 
This density corresponds to a Wigner-Seitz radius $r_{ws}$ = 2.255 a.u. for the ions
 and $r_s$ = 1.564 a.u. for the electrons since the mean
 ionization $\bar{Z}$ is found to be 3 from
the NPA calculation. They used two 4.5 J laser beams on both sides of a 50
$\mu$m-thick Al foil coated with a 2 $\mu$m-thick layer of Parylene. A
probe-pulse delay of $\tau_d =$ 1.9 ns is used. Hence the assumption of
thermal  equilibrium ($T_e=T_i$) seems justified. 
The NPA free-electron charge density $n_f(r)$ at an Al$^{3+}$
ion in the WDM system directly provides the pseudopotential $U_{ei}$ and the
pair-potential $V^\text{NPA}_{ii}(r)$. For the Yukawa screening of the YSRR potential, 
 Fletcher {\it et al.}  used the zero-$T$
value of the Thomas-Fermi wavevector (Eq.~\ref{ktf.eq}).
We find the
value of $\sigma$ to be 4.9 a.u., correcting what may be an error in 
Ref.~\cite{Flet-Al-15} where $\sigma$= 9.4 a.u.\ is quoted~\cite{sigCorr}. The
$S(k)$ corresponding to the NPA or the YSRR pair-potential 
can be calculated using an HNC or MHNC procedure, as appropriate, and
used in Eq.~\ref{xrts-eqn} to compute the XRTS-signal $W(k)$. 
 
\begin{figure}[t]
\includegraphics[width=.95\columnwidth]{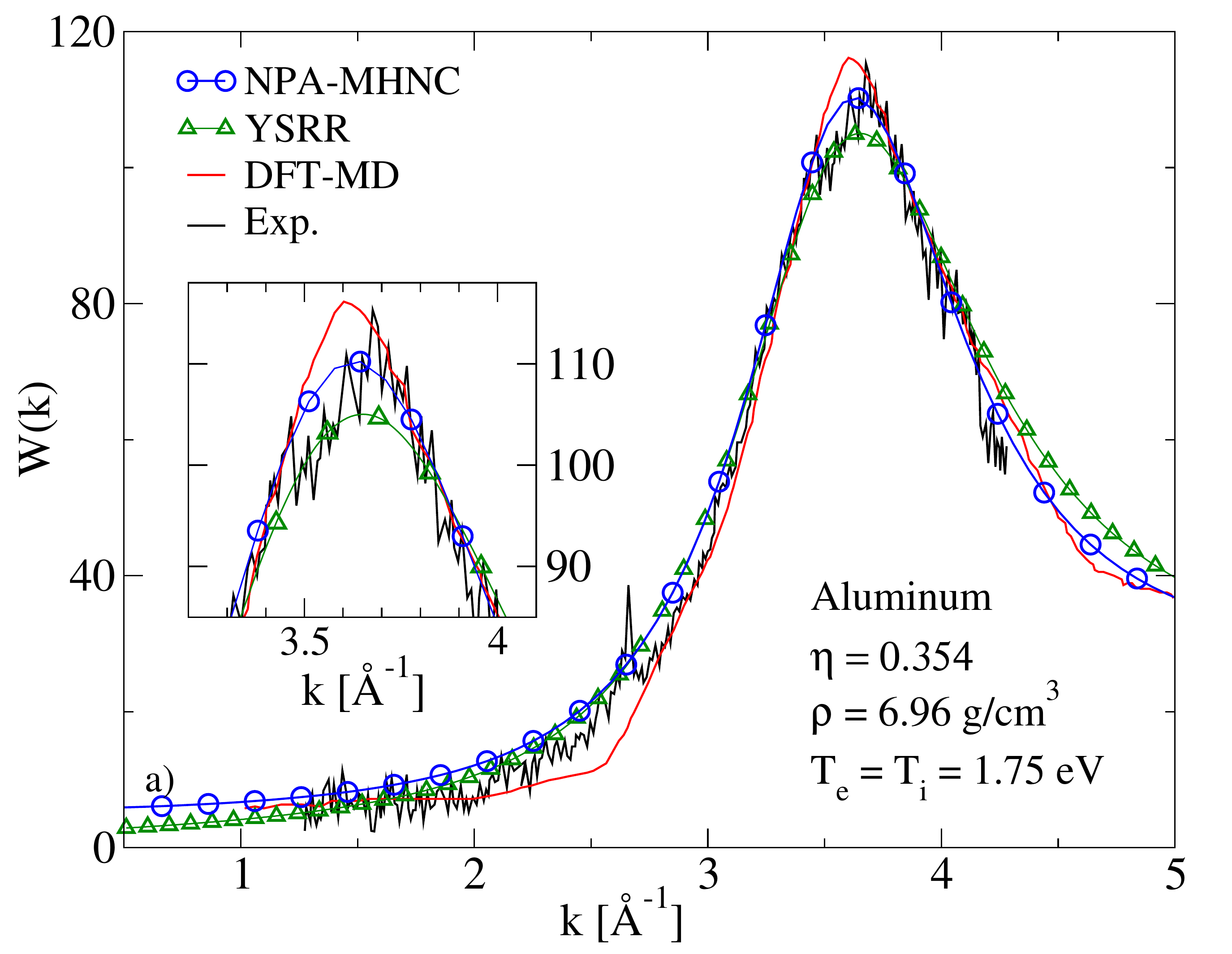}
\caption{The XRTS ion feature $W(k)$ of Ref.~\cite{Flet-Al-15} for Al-I,
 and from the DFT-MD, NPA-MHNC and YSRR models, as indicated. 
The inset magnifies the peak region.
}
\label{fig:WkFletcher}
\end{figure}

In Fig.~\ref{fig:WkFletcher}, the $W(k)$ computed from NPA and YSRR  are 
compared with the experimental XRTS $W(k)$. In the NPA case, a bridge
function $B(\eta,r)$ with $\eta=0.354$ is obtained from the LFA criterion.
The NPA-MHNC $W(k)$ is in good agreement with experiment and also
confirms thermal equilibrium with $T_i = 1.75$ eV. No bridge correction
is used for the YSRR since its $S(k\to 0)$ limit is already strongly inconsistent
with the compressibility sum rule as will be illustrated below in the discussion
(section~\ref{disc-section}).
%
%
\begin{figure}[t]
\includegraphics[width=0.95\columnwidth]{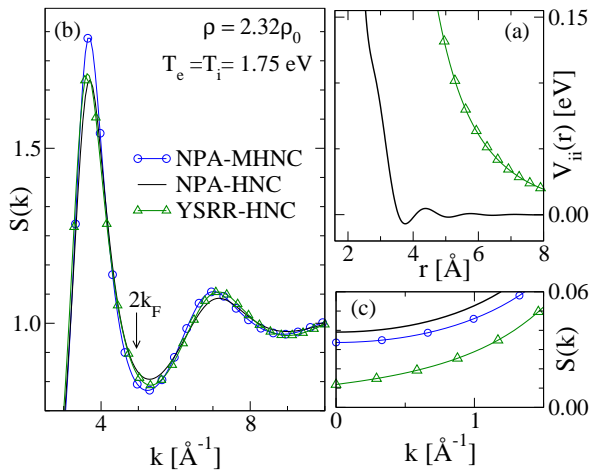}
\caption{(a) NPA and YSRR potentials for Al-I 
(cf. experiment of Ref.~\cite{Flet-Al-15}); 
(b) $S(k)$ from the $V_{ii}(r)$ using
 HNC and MHNC; (c) $k\to 0$ region of $S(k)$.
}
\label{fig:vr_sk_fletcher}
\end{figure} 
Since the conditions of the Fletcher experiment produce a  near-degenerate
 electron gas ($T_e/E_F=0.085$), the pair
potential $V^{\text{NPA}}_{ii}(r)$  displays Friedel oscillations as can be
seen in  Fig.~\ref{fig:vr_sk_fletcher}(a). The $S(k)$ from
NPA-HNC  and YSRR-HNC  are shown in Fig.~\ref{fig:vr_sk_fletcher}(b). The
NPA-HNC $S(k)$ is very similar to the YSRR-$S(k)$  but differs in the $k\to 0$
region  and near 2$k_F$ (panel (c)), and this will affect some EOS properties.  
\subsection{Shock-compressed Aluminum-II}
\begin{figure}[h]
\includegraphics[width=1.1\columnwidth]{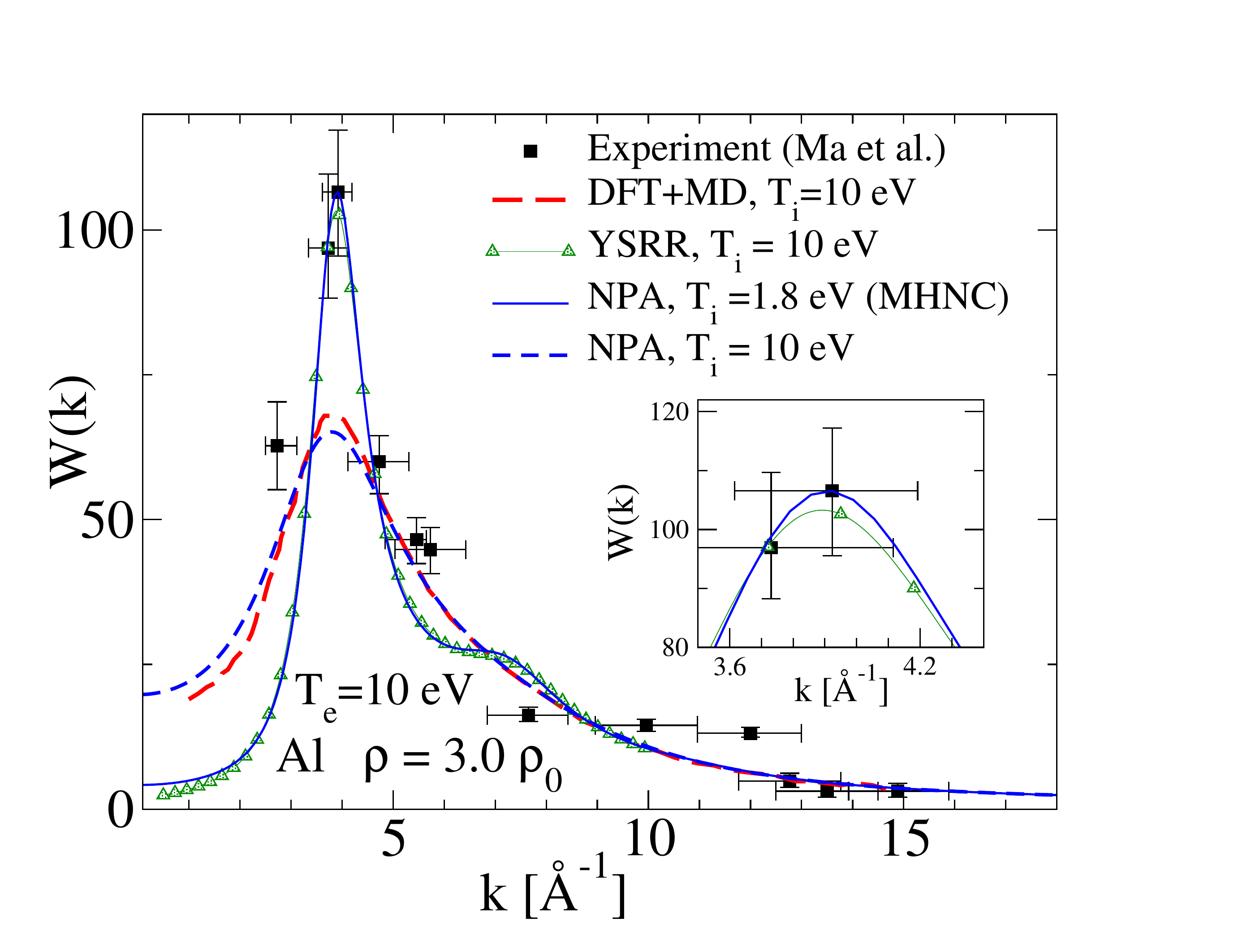}
\caption{The XRTS ion feature $W(k)$ of Ma {\em et al.}~\cite{Ma-Al-13} for Al-II,
compared with the $W(k)$ from YSRR, NPA, the DFT-MD calculation of
R\"{u}tter \textit{et al.} with $T_i=T_e$, and our $2T$-NPA
 calculation with $T_i=1.8$ eV and
$T_e=10$ eV; the inset magnifies the peak region.} \label{fig:WkMa}
\end{figure}

From inelastic data for shock-compressed Al (hereafter Al-II), Ma \textit{et
al.}~\cite{Ma-Al-13} determined the experimental conditions in the target to be
$\rho/\rho_0 = 3.0$ and $T_e = 10$ eV.
  This density corresponds to
$r_{ws}$ = 2.07 a.u. for the ions and $r_s$=1.435 a.u. for the
electrons. A set of nine pump beams, with a total energy of 4.5 kJ
 deposited in 1 ns, were
aimed directly at the 125 $\mu$m-thick Al foil without any protective shield. 
The shock compression heats up the ion subsystem on the
picosecond timescale, but the  coupling of the laser to the free electrons in
aluminum raises the electron temperature much more rapidly, within femtoseconds,
creating  a $2T$-system with $T_i<T_e$ initially.  If the data are collected 
after a sufficient time delay, an equilibrium temperature $T=T_i=T_e$ will  be
reached. Calculations using the  YSRR potential  with $T_i=T_e=10$ eV   show
good agreement with the XRTS ion feature. However, this turns out to be misleading
since the ion feature of the system at $T_i = 10$ eV determined by the
DFT-MD  simulation of R\"{u}ter \textit{et al.}\cite{Ruter}  disagrees with  the 
XRTS data of Ma \textit{et al.} as shown at Fig.~\ref{fig:WkMa}.  

\begin{figure}[h]
\includegraphics[width=\columnwidth]{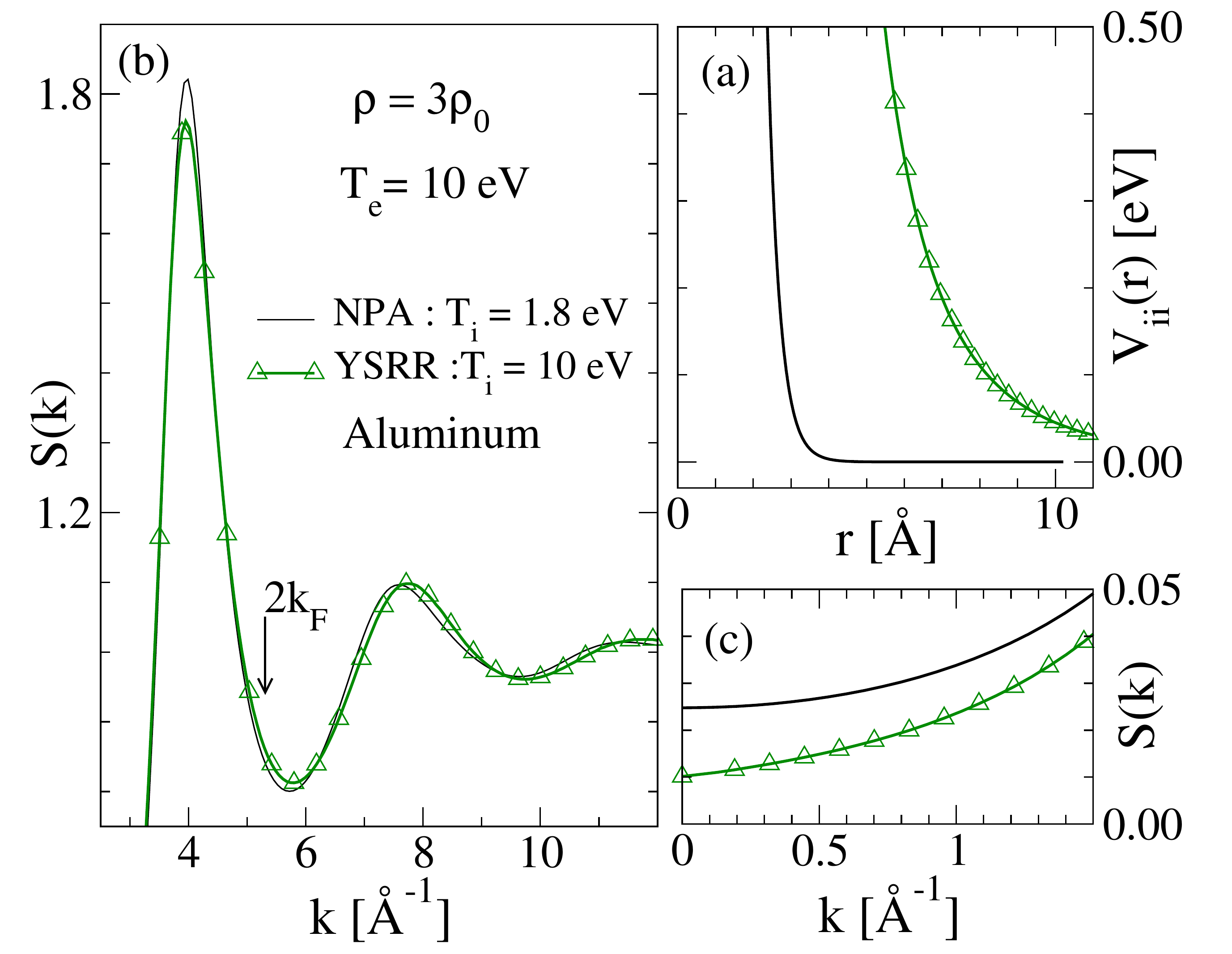}
\caption{(a) The NPA-HNC and YSRR  pair potentials for Al-II;
 (b) corresponding  $S(k)$ ; (c) the $k\to 0$ limit of $S(k)$.} 
\label{fig:vr_sk_ma}
\end{figure}

Using an ``orbital-free'' approach, viz.\ a Thomas-Fermi model with
Weisz\"{a}cker  corrections, Cl\'{e}rouin {\it et al.} \cite{Clerouin} arrived
 at a $2T$ model with $T_i = 2$ eV and $T_e=10$ eV in order to obtain
 good agreement with the XRTS data. They claimed that,
since their method involves all electrons, the core-repulsion term included
 in the YSRR model is non-physical. Our NPA Kohn-Sham calculations --- which are
 all-electron, 
parameter-free and include core and continuum states ---
confirm the conclusions of Cl\'{e}rouin {\em et al.}. 
Using the NPA-potential for this case, a MHNC calculation with $\eta=0.367$
predicts an excellent fit to the Ma {\em et al.}
 data with $T_i =1.8$ eV and $T_e = 10$ eV, as
can be seen in Fig.~\ref{fig:WkMa}.  In Fig.~\ref{fig:vr_sk_ma} (a)-(c), the NPA and
YSRR $S(k)$, pair-potentials $V_{ii}(r)$ and the $k\to 0$ limit of $S(k)$ are
shown.  There are no Friedel oscillations in  $V^{\text{NPA}}_{ii}(r)$ as $T_e$
is nearly six times higher  than in the conditions prevailing in Al-I. The
 disagreement between the NPA-$S(k)$ and the YSRR-$S(k)$ for
 $k\to 0$ should again be noted.
 
The high-$k$ shoulders of  the $W(k)$ curves from the $2T$ NPA-HNC
 and from the YSRR calculations are washed-out in the experiment, suggesting 
more complexity than in a $2T$ system. The ion subsystem  may be cold (at 1.8
eV), but containing an unknown high-$T$ component as well. On the other hand, it
has been pointed out by Souza \textit{et al.}~\cite{Sandia1} that the high
intensity peak around $k\sim 4 \AA^{-1}$ might be anomalous and caused by a
non-Gaussian and/or broadened probe beam.   The DFT-MD as well as the NPA results
for the equilibrium case ($T_e=T_i=10$ eV) both predict a peak height of
 $\sim 65$, in strong contrast to the YSRR
model, while the actual experimental peak height is $\sim 106$.

Evidently, the XRTS data cannot be consistent with an equilibrium model. Since
the aluminum target is pumped directly with a laser, the system would initially
begin with $T_e>T_i$, and the possibility that the system has
$T_e=10$ eV, with the ion subsystem  at $T_i\sim 2$ eV, is an entirely reasonable
result. More complex non-equilibrium features~\cite{ChapGerike2011} may also be
envisaged, and may be useful for explaining the wings of the XRTS data. A model
of the hot electrons involving a high-energy tail, energy bumps etc.,
 would involve additional parameters that fit the observed $W(k)$, but without
independent information to confirm them. Since the main XRTS $W(k)$ profile can be
explained well within a $2T$-model, this WDM is best regarded as being
 in a state with cold ions and hot electrons,but this by no means 
  excludes more complicated situations which can  be assessed only if more details
  of the experimental configuration and the pulsed heating process are
  available.

\subsection{Temperature relaxation in Al-II}
In this section, our objective is to estimate the temperature relaxation time
 $\tau_{ei}$ in order to determine if the system has reached equilibrium
 within the pulse-probe delay. Assuming the $2T$ model with $T_e=10$ eV
 and $T_i=1.8$ eV, we can use the NPA
electron-ion pseudopotential
to calculate the temperature relaxation rate~\cite{elr2001} via the
 Fermi Golden Rule (FGR). 
Since the FGR estimate of $\tau_{ei}$ 
was already sufficiently informative, we did  not need a more detailed
 energy-relaxation model such as the coupled-mode description which is
 known to  make the relaxation-time estimates somewhat longer. For
 this purpose we define the form factor $M_k$ of the pseudopotential
by
\be
M_k=U_{ei}(k)/\bar{Z}(Z_eV_k);\quad Z_e=-1.
\ee
The energy-relaxation rate, i.e., the rate of transfer of energy from
the hot electrons to the ions per unit volume,  
calculated within a number of
simplifying approximations, is given by
\begin{align}
\notag \frac{dE_e}{dt} &=  \omega _{i,\text{p}}^{2}( T_{e}-T_{i}) \\
&\times \int \frac{
d^{3}\vec{k}}{\left( 2\pi \right) ^{3}}\ V_k
\left| M_{\vec{k}}\right| ^{2}\text{Im}\left[ \frac{\partial }{\partial
\omega }\chi^{ee}(\vec{k},\omega) \right] _{\omega =0},
\label{trr1}
\end{align}
where $E_e$ is the energy in the hot electron subsystem at time $t$,  
$\omega _{i,\text{p}}$ is the ion plasma frequency and $\chi^{ee}$ is 
the fully interacting dynamic electron response function. The $f$-sum
 rule has been used to eliminate the dependence on the 
ionic structure~\cite{Hazak} and hence provides an estimate of  the
energy relaxation which is superior to those obtained by using models for
$S_{ii}(k,\omega)$,  if the other assumptions made in the above theory hold.  For
instance, a difference of Bose factors of the form $N(\omega/T_e)-N(\omega/T_i)$,
where $N(x)=1/(\exp(x)-1)$, for the density fluctuations in the
electron system and the ion system respectively, has been approximated as
$(T_e-T_i)/\omega$ in order to calculate a temperature relaxation rate. If $T_i$ is
assumed to be fixed (at least for a short timescale), we can use Eq.~\ref{trr1}
 for the energy relaxation rate
of the electrons to determine a temperature relaxation rate. It requires a
 relation between
the internal energy of an interacting warm-dense electron fluid and its
temperature. Here we use the property that $F=F_0+F_{xc}$, and the internal
energy $E=\partial \{\beta F\}/\partial \beta$ as presented in Ref.~\cite{PDWXC}
where the needed parametrizations are given. The replacement of the Bose factors
by a temperature difference is not quite valid for the $T_i=1.8$ eV and $T_e=10$
eV estimated to prevail in the Ma {\it et al.} experiment since the electrons are
partially degenerate. 

 Nevertheless, one can obtain a {\it grosso modo} estimate
of the temperature-relaxation time $\tau_{ei}$. It is found to be 300-400 ps
depending on various assumptions. This timescale is sufficient
for the formation of phonons, and hence the temperature relaxation towards
equilibrium will be further slowed down by the formation of coupled modes (i.e.,
the conversion of ion-density fluctuations by electron screening into
ion acoustic modes). This slows down the relaxation time by more than an order of
magnitude. An actual estimate of the temperature relaxation of the target
material (Al-II)  will also have to account for the fact that the ion subsystem
looses energy to its holding structure and the environment. These considerations
independently support our conclusion, and that of Cl\'erouin {\it et al.},
that $T_i < T_e$ is a possible scenario, contrary to the `equilibrium model'
 indicated by the YSRR model. 
\subsection{Discussion of Al results}
\label{disc-section}
\label{disc-Al-I}
DFT simulations treat the WDM sample as a periodic crystal made up of $N$ nuclei
whose positions in the simulation box evolve via MD and provide
``single-electron'' Kohn-Sham spectra. However, it provides no simple method for
computing electron properties that can be attributed to a ``single'' nucleus, e.g.,
the mean ionization $\bar{Z}$ arising from the bound and free
parts of the spectrum, or  pair interactions resulting from the single ions.
The latter, if available,
provides a convenient means of obtaining $S(k)$ and related properties of the WDM
in a computationally efficient manner. The YSRR model potential was justified by
Wunsch \textit{et al.}~\cite{Wunsch09} as a suitable way of inverting a given
$g(r)$ obtained from DFT-MD simulations in WDM conditions; it contains a
Yukawa-like screening term based on an externally provided $\bar{Z}$ and an 
explicit ``core-repulsion" term.

The NPA approach rigorously constructs the effective Kohn-Sham
``single-electron'' density via the ``single-ion''  DFT description of
 the electron-ion system, as
implied by the Euler-Lagrange equations given by Eq.~\ref{ks-el.eq}
and Eq.~\ref{ion-dft.eq}. Thus, unlike DFT codes which treat the ions as an external
potential, the NPA directly provides single-ion/single-electron properties as
well as the pair potentials. The NPA calculation for Al-I, for the experiment of
Fletcher {\it et al.}, shows that the mean radius of the $n=2,l=0$ bound shell in
Al, which reflects the radius of the bound core, is 0.3552 \AA. The YSRR
potential reaches large values already by 2 \AA, i.e., at a radius nearly 6 times
larger than the actual core size; thus the short-range repulsive part
$(\sigma/r)^4$ is not appropriate.  The claim in
Ref.~\cite{Flet-Al-15,Ma-Al-13} that the YSRR potential ``accounts for the
additional repulsion from overlapping bound-electron wavefunctions''  is
certainly not confirmed by the shell structure of Al$^{3+}$ in the plasma.  Note
that even the Wigner-Seitz radius, i.e., the sphere radius for an ion for
aluminum at a compression of 2.32, is 2.255 a.u.$=$ 1.193 \AA$\,$,  and hence the
YSRR model is clearly unphysical. The core-core interaction in Al can be 
calculated from the Al$^{3+}$ core-charge density as in Appendix B of
Ref.~\cite{eos95}. It is totally negligible for Al at compressions of 2.32 (in
Al-I) or 3 (in Al-II) studied here. It should be noted that, as far as $S(k)$ is
concerned, core-core interaction effects lead to an {\it attraction} due to
core polarization, as was also discussed in 
Ref.~\cite{eos95}. This too can be neglected in aluminum.

The liquid-metal  community of the 1980s found  that the inverse problem of
extracting a potential from the  $S(k)$ given in a limited $k$-range,  obtained
from MD or from experiment, is misleading and  not
unique~\cite{March,AersCDW,Rosenfeld}. A parametrized   physically-valid model (e.g., a
pair potential $V_{ii}(r)$ constrained via an atomic pseudopotential) together
with a good $B(\eta,r)$~\cite{chenlai} can successfully invert the MD data.
However, the  DFT-MD step is unnecessary in most cases since the
$V^{\text{NPA}}_{ii}(r)$  and the $S(k)$ that provide the  physics are easily
evaluated from a rapid parameter-free NPA calculation. The YSSR potential is fitted to
a limited range in $r$-space as in Wunsch {\it et al}$\,$.
 But the Fourier transform
 to obtain $k$- space
quantities involves information on all of $r$-space. This
leads to serious and uncontrolled errors unless a physically valid potential is
used to extend the simulation $g(r)$ data to all $r$ and hence to all
 relevant $k$.
We note the following.
 (a) The YSSR is proposed in  Fletcher {\it et al.} for the
 computation of the EOS. Small-$k$ behaviour is very
 important for EOS properties and we bring this out via the compressibility
 calculation. (b) Behaviour near 2$k_F$ is important for transport and scattering
 processes and we bring this out via the resistivity calculation. (c) Other
intermediate $k$-values are sampled by the phonons and we show that  YSSR
fails for most-$k$ in the phonon dispersion.
 Hence  even when YSSR `seems to work' for one property, one cannot attribute
 any  physical significance to it. Thus, besides the XRTS ion feature, we tested the validity of NPA and YSRR models under WDM conditions by computing three key physical quantities, namely (i)
compressibility $\kappa$, (ii) phonon spectra, and (iii) resistivity $R$,
 as we discuss below. 

(i) To determine $\kappa$ we assume that the sum rule $S(0)=\rho T_i \kappa$ holds for
$2T$ systems under certain restrictions~\cite{Bonarth}.  We computed $\kappa$
using NPA, YSRR and ABINIT, and obtained respectively 26, 9.6 and 30 a.u. for Al-I
at a compression of 2.32 and $T_i=T_e=1.75$ eV. The corresponding values for
aluminum (Al-II) at a compression of 3, $T_i=1.8$ eV and $T_e=10$ eV, are  14, 1.1
and 16.4 a.u. In both cases, the results from the NPA are in close agreement
with ABINIT whereas the YSRR gives a much lower compressibility. Here, the YSRR-$S(k)$ is
calculated from the HNC without a bridge term, i.e. $B(\eta,r)=0$, since a
bridge term would make the compressibility even more erroneous. Thus, even in
equilibrium, the YSRR model is not trustworthy enough for EOS properties like the
compressibility. 

\begin{figure}[h]
\includegraphics[width=\columnwidth]{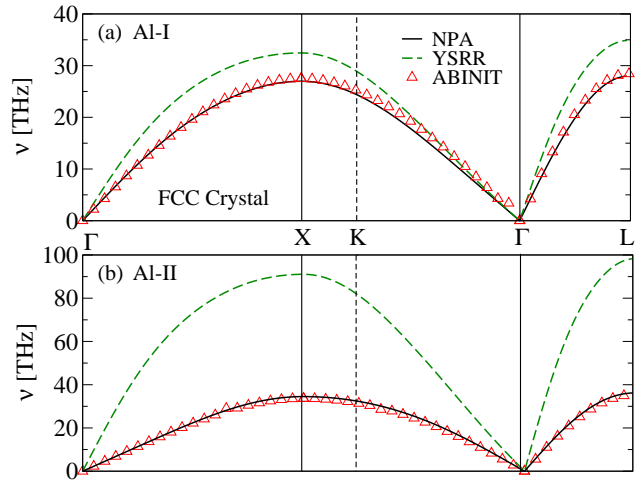}
\caption{The longitudinal phonon spectrum for the FCC crystal of a) Al-I, i.e.
compression of 2.32 and $T_e=T_i=$1.75 eV, 
Ref.~\cite{Flet-Al-15}; b) Al-II, i.e.  compression of 3.0,
 $T_i$=1.8 eV, and $T_e=10$ eV,  Ref.~\cite{Ma-Al-13}. The $\Gamma$, X, K and L point are symmetry points of the first Brillouin zone of the FCC crystal.} 
\label{fig:phon_fe_ma}
\end{figure}

(ii) Even though the ionic system is clearly melted in both Al-I and Al-II conditions, a good test of the quality of the pair potential is the computation of  the phonon spectrum for its low-$T$  crystal structure  (face-centered-cubic (FCC) for Al) which is a particular ionic configuration of the system even in the melt. 
In fact, the short-range structure of strongly coupled ionic fluids as reflected in the $S(k)$ is known to correspond closely to the $S(k)$ of the crystal structure below the melting point. The comparison of phonons obtained via the pair potential approach with those from \textit{ab initio} calculations  permits the validation of the energy landscape created by the pair potential for this particular ionic configuration.  Such tests have already been done for other systems showing that the NPA  predicts equilibrium and non-equilibrium phonons in good accord with
\textit{ab initio} simulations ~\cite{CPP-Harb}, which illustrates its meV-level of accuracy. The examination
 of phonon modes is relevant for ultra-fast-matter (UFM) studies where electron temperatures will rise significantly more
 rapidly than that of the nuclei. The limiting case where the nuclei are at low temperature is where
 phonon stability is relevant The excellent agreement between the NPA and ABINIT longitudinal phonons in Al-I (Ref.~\cite{Flet-Al-15}) and Al-II (Ref.~\cite{Ma-Al-13}) is displayed in Fig~\ref{fig:phon_fe_ma} and further validates the NPA pair potentials in the WDM regime. The unphysical
``stiffness'' of the YSRR potential leads to high phonon frequencies and a sound
velocity much larger than  the NPA and ABINIT predictions. 

\begin{figure}[h]
\includegraphics[width=1\columnwidth]{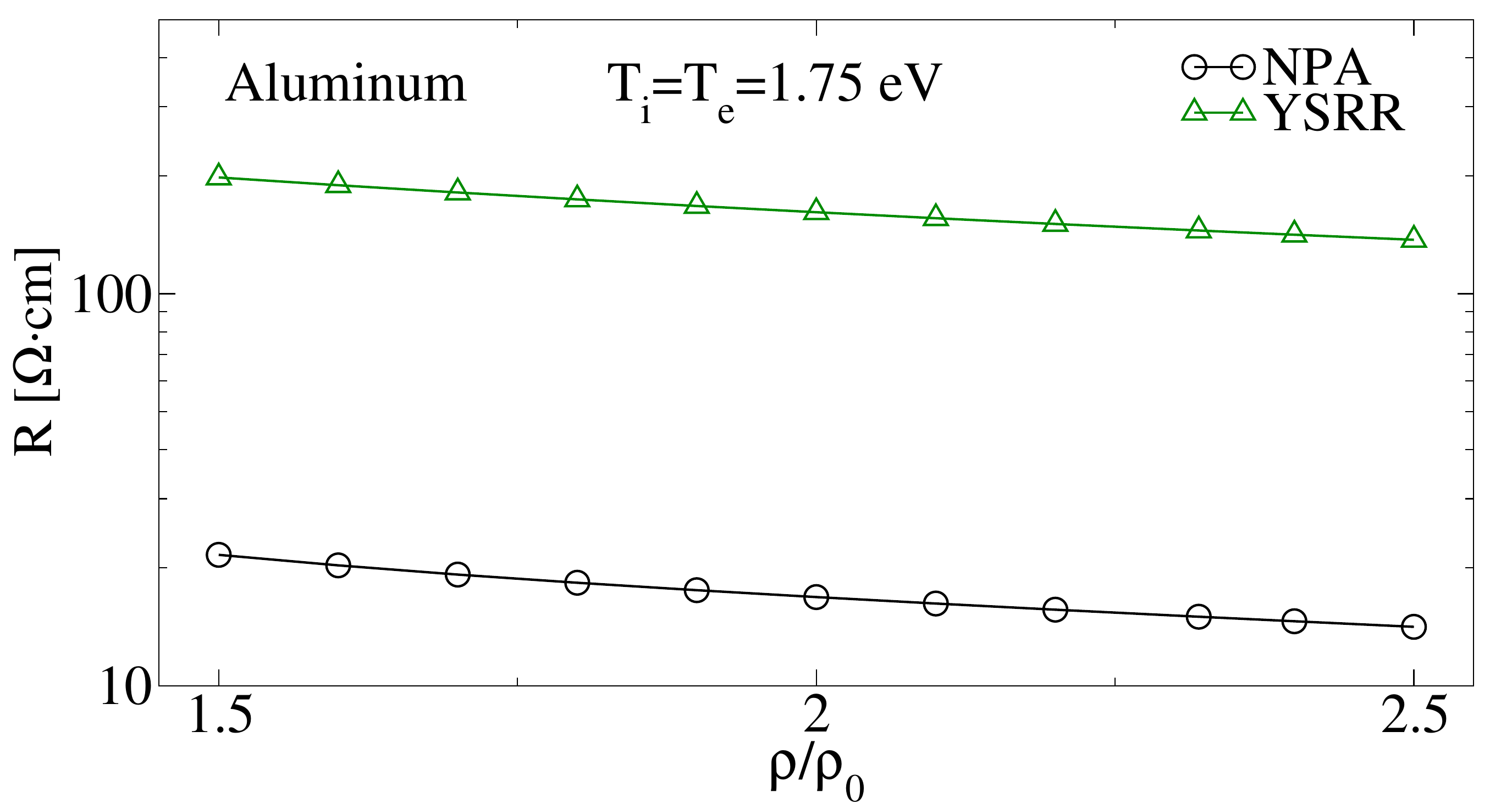}
\caption{The electrical resistivity of Al at T=1.75 eV for different compressions
 calculated using NPA and YSRR.
} 
\label{fig:yssr-res}
\end{figure}

(iii) We tested the validity of the Yukawa component  in the YSRR model and the
validity  of the YSRR-$S(k)$ by calculating the electrical resistivity $R$. The
Yukawa pair potential  $\bar{Z}^2 \exp(-k_s r)/r$ arises from  the Yukawa
pseudopotential $U^y_{ei}(q)=-4\pi \bar{Z}/q^2$ screened by the $k\to 0$  RPA
dielectric function, i.e., $\epsilon(q)=1+(k_s/q)^2$.  We use the Ziman formula
in the form given in Ref.~\cite{eos95},~Eq.~(31) to calculate the resistivity.
Computing $R$ for the NPA and YSRR model, we
obtain respectively  15.0 and 145 $\mu\Omega\cdot $cm for  Al-I while the
 corresponding values for Al-II are  9.65 and  99.4  $\mu\Omega\cdot $cm. Thus,
 in both cases, the resistivity predicted
 by the YSRR is about 10 times higher than the NPA value. Such
 larger-than-expected resistivities have also been obtain by Sperling
 \textit{et al.}\cite{Sper15} while using an even simpler model than YSRR.
The resistivities predicted by Sperling {\it et al.} are known to be in strong
disagreement with the DFT-MD Kubo-Greenwood resistivity calculations of
Sjostrom {\it et al.}~\cite{SjosCond2015}. These issues are discussed at length
 in Ref.~\cite{cdwPlasmon} where it is concluded that the Sperling calculation
of the static conductivity is likely to be inapplicable.
 In Fig.\ref{fig:yssr-res},
 we show that this behavior is also observed for various densities in equilibrium
  with $T = 1.75$ eV. The Ziman formula in conjunction with the NPA-HNC model,
 which includes a self-consistently generated $U_{ei}(k)$, $S(k)$
and a screening function $\chi(k)$ containing an LFC that satisfies the
compressibility sum rule, is a well-tested method for many systems 
(for a review, see Ref.~\cite{CPP-cdw}) including
aluminum~\cite{eos95,Benage99}. Thus, while it may be thought that
additional \textit{ab initio} or experimental resistivity data are required
 to confirm the NPA and Ziman formula results in the WDM regime, it is unlikely
that the NPA resistivities are in error by an order of magnitude, given
 the excellent track record of NPA-resistivity predictions~\cite{Benage99}.
In our view, the Yukawa part of the YSSR calculation is responsible for the
erroneous estimate of the resistivity.

In summary, through the calculation of phonons,
 compressibilities and resistivities, 
we  showed that the short-ranged-repulsive and the Yukawa
 parts of the YSRR model are individually untenable. In contrast,
the correct physics is quite simply obtained from the NPA-HNC for both
 the equilibrium and the $2T$ situation. 

\section{Beryllium}
\subsection{Shock-compressed Beryllium}
\begin{figure}[h]
\includegraphics[width=0.9\columnwidth]{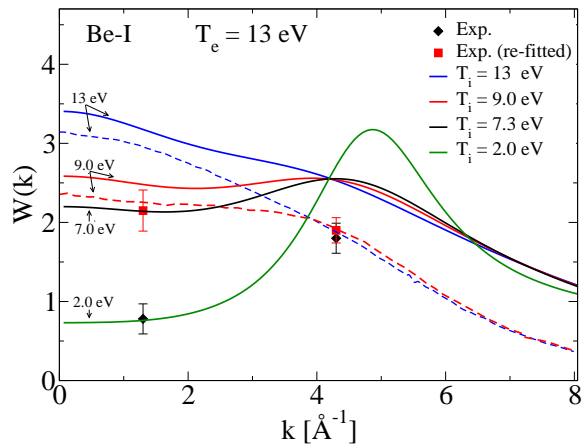}
\caption{The XRTS  ion feature $W(k)$ of Lee \textit{et al.}~\cite{Lee-Be-09}
compared with the NPA-HNC $W(k)$ (full lines) and with  DFT-MD
results (dashed lines) of Plagemann \textit{et al}~\cite{Plag15} for 
equilibrium and for $T_e \neq T_i$, as indicated.
} 
\label{fig:Be1}
\end{figure}

\begin{figure}[h]
\includegraphics[width=0.9\columnwidth]{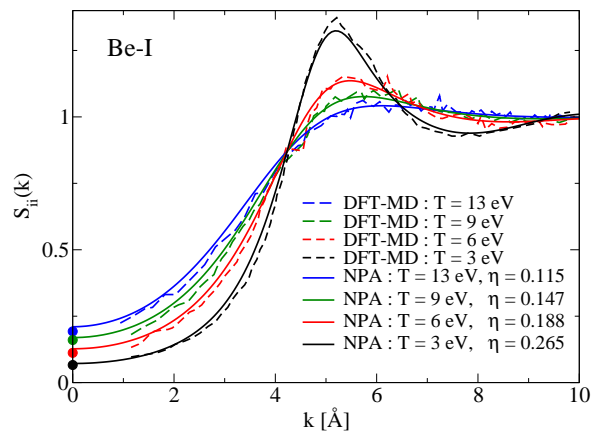}
\caption{Static ion-ion structure factor $S(k)$ for Be-I from the NPA-HNC and
DFT-MD simulation of Plagemann \textit{et al}~\cite{Plag15}, including their
$k=0$ value marked as dots, for different equilibrium temperatures.} 
\label{fig:Be2}
\end{figure}

\begin{figure}[h]
\includegraphics[width=0.9\columnwidth]{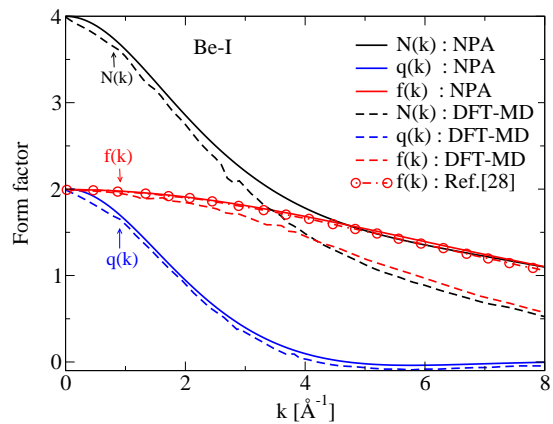}
\caption{Total $N(k)$, bound-electron $f(q)$ and free-electron
 $q(f)$ form factor for the
Be-I conditions} 
\label{fig:Be3}
\end{figure}

Beryllium has been of recent interest (e.g., Ref.~\cite{DafangLi14}) 
 for many reasons including its potential
applications as an ablator material in inertial-confinement fusion studies.
Lee \textit{et al.}~\cite{Lee-Be-09} studied compressed beryllium
  by applying 12 pump beams,
  each with an individual
energy of 480 J in 1 ns, directly on a 250 $\mu$m-thick Be foil without any
coating. The pump-probe laser delay  is $\sim$ 4.5 ns, and may appear
to be enough for electron-ion equilibration. We will examine this by a
calculation of the $\tau_{ei}$, as was done for aluminum. From an analysis of the
XRTS data they concluded that Be is in a compressed state with $\rho/\rho_0
=2.99$ and $T_e = 13$ eV (Be-I hereafter). Figure~\ref{fig:Be1} compares
 the ion feature $W(k)$ from
the NPA-HNC model  with the experimental data of Ref.~\cite{Lee-Be-09} and
with the detailed and careful DFT-MD simulations of
 Plagemann \textit{et al.}~\cite{Plag15}.
Even though NPA-HNC and DFT-MD do not predict exactly the same spectrum,
both approaches agree in not
confirming the first experimental point at $k = 1.3\ \AA^{-1}$
 under the equilibrium
condition $T_i=T_e=13$ eV.  By re-analyzing the original data
 (indicate as  `Exp (re-fitted)' 
in Fig.\ \ref{fig:Be1} and in Fig.\ \ref{fig:Be4}), Plagemann \textit{et al.}
found that a $2T$-system with $T_i = 9$ eV and $T_e = 13$ eV was able to
reproduce the spectrum. The NPA-HNC calculation does not predict the second
experimental point at $k = 4.3\ \AA^{-1}$ as it shows higher values than DFT-MD
for all values of $k$. To understand this difference between DFT-MD and NPA-HNC,
we compared individually the two contributions to the
ion feature, namely the static ion-ion structure factor $S(k)$ and the total
electron form factor $N(k) = f(k) +q(k)$.
In Fig.~\ref{fig:Be2}, a comparison of
the $S(k)$ at different equilibrium temperatures shows excellent accord
 between NPA-HNC and DFT-MD.
However, Fig.~\ref{fig:Be3} reveals important differences in the core-electron
 form factor $f(k)$ starting around $k = 4\ \AA^{-1}$. To determine
$f(k)$, Plagemann \textit{et al.} used snapshots of the DFT-MD simulation
from the VASP code and post-processed in the ABINIT code using a
 ``superhard'' pseudopotential accounting for all four electrons. In contrast, 
NPA-HNC is an all-electron atomic calculation including corrections
from the ion environment which predicts an $f(k)$ similar to the independent
pseudo-atom calculation of Souza \textit{et al.}~\cite{Sandia1}.  Since the
discrepancy starts around  $k = 4 \AA^{-1}$, i.e., already deep into the
core, it is possible that the  ``superhard'' pseudopotential used
does not reconstruct correctly 
 the core electron density close to the nucleus. 
Further investigation should be done of this possibility which
 would account for the
differences in the calculation of $W(k)$ from NPA-HNC and DFT-MD.
 Finally, using the NPA-HNC
model while keeping $T_e = 13$ eV, the best fit to the re-analyzed
experimental $W(k)$ is obtained with $T_i = 7.3$ eV  while it requires $T_i =
2$ eV to reproduce the original data. Given $T_i=7$ eV and $T_e=13$ eV, the
Be-target is better equilibrated than if one were to posit $T_i=2$ eV,
and $T_e=13$ eV.
\subsection{Isochorically-heated Beryllium}
\begin{figure}[h]
\includegraphics[width=0.9\columnwidth]{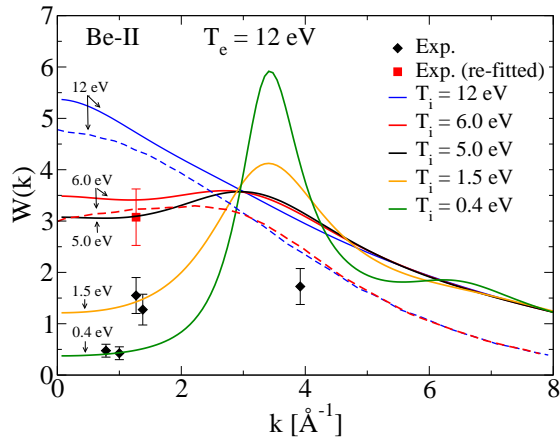}
\caption{The XRTS ion feature $W(k)$ of Glenzer 
\textit{et al}~\cite{Glenzer-Be-07} compared with the $W(k)$
 of the NPA-HNC
 model (full lines)
and with the DFT-MD simulation (dashed lines) of 
Plagemann \textit{et al}~\cite{Plag15} for the equilibrium and $2T$ situations.
} 
\label{fig:Be4}
\end{figure}

\begin{figure}[h]
 \vspace{0.2cm}
\includegraphics[width=0.9\columnwidth]{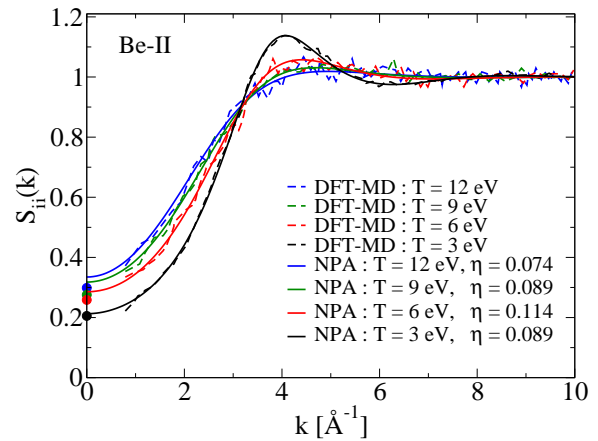}
\caption{Static ion-ion structure factor $S(k)$ for
 Be-II from the NPA-HNC model and
from the DFT-MD simulations of Plagemann \textit{et al}~\cite{Plag15},
 including the
$k=0$ values marked as dots, for different equilibrium temperatures.} 
\label{fig:Be5}
\end{figure}

Glenzer \textit{et al.}~\cite{Glenzer-Be-07} created an
 isochorically heated ($\rho/\rho_0=1$)
WDM Be sample (named Be-II hereafter) by aiming 20 pump beams, with a
total energy of 10 kJ over 1 ns, onto a 300 $\mu$m-thick Be cylinder coated by a
protective 1 $\mu$m-thick silver layer. They determined that
 $T_e = 12$ eV while the pump-probe delay
of 0.5 ns was considered sufficient to achieve thermal equilibrium
 between ions and electrons. In
Fig.~\ref{fig:Be4}, the $W(k)$ from the NPA-HNC model is compared with the
experimental data  of Ref.~\cite{Glenzer-Be-07} and with the DFT-MD simulations
of Plagemann \textit{et al.}~\cite{Plag15}.
The NPA-HNC and the DFT-MD calculations do not reproduce the original
 experimental
data of Glenzer {\it et al.} nor the re-analyzed data using equilibrium
conditions with $T_i=T_e=12$ eV.
By keeping $T_e = 12$ eV, Plagemann \textit{et al.} found that it was possible to
reproduce their (single) re-analyzed data point by setting $T_i = 6$ eV. In order to
reproduce this point, the NPA-HNC model requires a slightly lower ion temperature
of $T_i= 5$ eV. Within the NPA, it was impossible to reproduce all four original
experimental points  for $k< 2 \AA^{-1}$ with a single $T_i$. However, the first
two points could be obtained with $T_i = 0.4$ eV while the two next points
required $T_i = 1.5$ eV. Both models are not able to reproduce the point at $k=3.9
\AA^{-1}$ since it  is too low to be reproduced with any $T_i$. 
As in the case of isochoric compressed Be studied
 by Lee {\it et al.}~\cite{Lee-Be-09},
the NPA-HNC model predicts  higher $W(k)$ values 
than DFT-MD simulations for all $k$. A comparison
between the NPA-HNC and the DFT-MD $S(k)$ is shown in Fig.~\ref{fig:Be5}, 
demonstrating close agreement between results from the NPA
pair potential and DFT-MD. Hence, in this case also the difference
 in $W(k)$ between the two
methods comes from the difference in the core electron form factor $f(k)$,
which is essentially similar to the compressed-Be case presented in
Fig.~\ref{fig:Be3}. Whether the ``superhard'' pseudopotential used
 by Plagemann \textit{et al.} 
 or other plasma effects included in the NPA treatment,
 but not in the DFT-MD, may be responsible for differences
 in the core electron density near the nucleus is unclear at present. 
We again note that the NPA is an ``all-electron'' method.
\subsection{Discussion of Be results}
In addition to the differences between NPA-HNC and DFT-MD in the $k> 4\ \AA^{-1}$
region for $W(k)$, we also observe a disparity at $k = 0$ for both
 Be systems studied here. Since the total electron form factor $N(k=0) = 4$ is equal
to the total number of electrons per ion, the divergence between the two models
comes from the structure factor $S(k=0)$ as shown in Figs.~\ref{fig:Be3} and
\ref{fig:Be5}. It is very difficult to reach such low $k$ values from
DFT-MD simulations (because of the finite size of the simulation cell)
 and Plagemann \textit{et al.} extracted them from independent
thermodynamic calculations. As mentioned before, this quantity is
important since it is linked to the compressibility $\kappa$ via the sum rule
$S(k=0)=\rho T_i \kappa$. It should be noted that an accurate
value of the compressibility $\kappa$ should be determined
 from an EOS calculation, while
the $S(k\to0)$ limit, from an MHNC calculation,  matches the EOS-$\kappa$ only
 when the bridge contribution is optimal. Table~\ref{kappa-table} compares
 calculations from the NPA-HNC method and from DFT-MD by Plagemann {\it et al.}.
\begin{table}
\caption{
Compressibilities $\kappa$ (in a.u.) of WDM beryllium calculated
 using DFT-MD and NPA-HNC.\\} 
\begin{ruledtabular}
\begin{tabular}{lcc}
 System        & Be-I (Ref.~\cite{Lee-Be-09})
 &Be-II (Ref.~\cite{Glenzer-Be-07}) \\
\hline \\
 $T_i\  [eV]$             & 13.0, 9.00, 6.00,  3.00& 12.0,  9.00,  6.00,  3.00  \\
\hline\\
$\kappa$ (DFT-MD)    & 3.72, 4.47, 4.69,  5.54& 18.6, 22.8, 32,1, 50.1 \\ 
$\kappa$ (NPA-HNC)  & 3.89, 4.73, 5.32, 6.01& 20.8, 26.3, 35.5, 52.8\\
\end{tabular}
\end{ruledtabular}
\label{kappa-table}
\end{table}
In both situations, the
NPA-HNC compressibility is slightly higher than the DFT-MD, which is enough to 
affect the small-$k$ region of the $W(k)$. This emphasizes the importance of
experimental data for $k = 0$ in order to validate theoretical models and
thermodynamics for WDM.
On the other hand, we saw that, in the Be-II case, the two models are unable
to predict the $k=3.9\ \AA^{-1}$ data and that it is impossible
 to reproduce all four
original experimental points for $k < 2\AA^{-1}$ using one single-ion
temperature. The extraction of $W(k)$ is highly dependent on how the other terms
in the Chihara decomposition, in particular the free-free electron-electron
structure factor $S_{ee}^0(k,\omega)$, are computed.   The $S_{ee}^0(k,\omega)$
is directly linked to the imaginary part of the response function $\chi$ and
most authors have been using the Mermin~\cite{Mermin-70} formulation while including
electron-ion collisions in the Born approximation. It has been recently
pointed out that the Mermin approximation is not applicable to UFM systems
because of inherent assumptions behind the model~\cite{cdwPlasmon}. Recently,
time-dependent-DFT-MD simulations have been done for Be~\cite{Baczewski-Be-16}
and it would be of interest to compare the total
electron-electron dynamic structure factor $S_{ee}(k,\omega)$, including the
bound-free contribution, in the WDM regime from these different models. 
Until a satisfactory model for 
$S_{ee}(k,\omega)$ is validated, it is hard to determine $T_i$ via $W(k)$ but it
should be kept in mind that a two-temperature system or other more complex
situations might occur in
laser-generated WDMs.
\subsection{Temperature relaxation in Be}
We consider the energy relaxation rate for the isochorically-heated beryllium
  (Be-II) where $\rho/\rho_0=1$, and 
for the particular $2T$-case with $T_e=12$ eV and $T_i=6$ eV. 
Here the electron-sphere radius $r_s \simeq 1.92$ and $T_e$ is
 close to the Fermi energy $E_F^{\text{Be-II}}= 16.6 $ eV. Hence
 this system is far less degenerate than
the Al-I and Al-II systems discussed above. The Fermi golden-rule calculation
using the $f$-sum rule gives a temperature relaxation time of 150-200 ps. 
Coupled-mode formation may slow this down by an order of magnitude. Hence the
claimed delay of about 500 ps may not be enough to achieve equilibration.
The difficulties in matching the experimental data with simulations
 also indicate
that we do not have a properly equilibrated  WDM-Be sample.

In the case of the compressed Be sample with $T_e=13$ eV (Be-I), 
 the $f$-sum-based relaxation time is nearly five times
faster than for Be-II. Hence temperature-equilibration shortcomings
cannot be an explanation for the difficulties encountered in modeling the
data using a $2T$ approach. Difficulties in reproducing the $W(k)$ using
   NPA-HNC and DFT-MD 
suggest that the experimental characterization requires further attention. 
\section{Conclusion}
We have presented parameter-free all-electron NPA-HNC calculations 
 of the charge densities, pseudo-potentials, pair potentials
 and structure factors
that are required to interpret XRTS experiments. Compressibilities, phonons,
 resistivities as well as temperature-relaxation times for relevant
 cases have been
presented, using the NPA pseudopotentials and structure factors where needed.
Re-analyzing recent WDM experiments enabled
 us to (a) investigate the validity of the  commonly-used YSRR model by  showing
that both its short-ranged part and its screening part yield
misleading predictions; (b) expose pitfalls in inverting structure data to
 obtain effective pair potentials;
(c) examine possible $2T$-models and their temperature relaxation
to examine the interpretations of $W(k)$ data from XRTS,  emphasizing the
need for caution in assuming thermal equilibrium 
in laser-generated WDM;  and  (d) demonstrate
the accuracy of the NPA calculations of physical properties of
 electron-ion systems,
 from ambient temperatures and  compressions to high temperatures
 and high compressions.
The computational rapidity of the NPA-HNC model permits `on-the-fly'
 testing-out of
possible values of $T_i, T_e$ and compressions that may rapidly fit an experiment,
 while this is time-consuming or impossible with DFT-MD simulations of
 properties like the ion feature $W(k)$ of WDM systems.
\section*{Acknowledgments} This work was supported by grants from the
 Natural Sciences and Engineering Research Council of Canada (NSERC)
 and the Fonds de Recherche du Qu\'{e}bec - Nature et Technologies (FRQ-NT).
 We are indebted to Calcul Qu\'{e}bec and Calcul Canada for generous
 allocations of computer
resources.


\begin{thebibliography}{[1]}
\bibitem{sigCorr}
We have contacted the authors of Ref.~\cite{Flet-Al-15} regarding this. If
$\sigma=9.4$, then the classical coupling constant at $r_\text{ws}=1$ a.u.
would be over 7800 at all temperatures as this term is independent of $T$. The
HNC does not converge for $\sigma\sim 9$, although it may be treated using MD.
\end{thebibliography}
\end{document}